# Diagonalizing the Hamiltonian of capacitively coupled superconducting phase qubits

[1]Tao Wu, [2]Zheng Li, [1]Jianshe Liu
[1]Institute of Microelectronics, Tsinghua University, Beijing, 100084, China
[2]Department of Electrical Engineering, Tsinghua University, Beijing, 100084, China

Coupling qubits together towards large-scale integration is a key point for realizing a quantum computer. We study the capacitively coupled superconducting phase qubits using two diagonalization methods, which are very efficient to obtain the wave functions and energies of the bound states of such two-qubit system. The first diagonalization method is based on two-dimensional cubic approximation for the coupled system with wave functions of the eigenstates for harmonic oscillators as the bases of diagonalization, and also reveals the physics underlying it. The other one utilizes the Fast Fourier Transform to perform diagonalization with plane waves as the bases, and it can be easily extended to other problems with even high dimension.
KEYWORDS: phase qubit, diagonalization, cubic approximation, FFT

## 1. Introduction

Superconducting phase qubits are current-biased Josephson junctions. [1-5) ] Compared with charge qubits and flux qubits made by Josephson junctions, [6-12)] their sizes are usually larger which make them more tolerable to the inaccuracy and asymmetry of technology processes; in addition, they are insensitive to both the charge and the flux noise in the background. Then it is extremely desirable to accomplish multi-qubit system utilizing phase qubits. Coupling two phase qubits through a capacitance is a basic one of many proposals and has already been investigated both theoretically and experimentally so far. [13, 15-19)] Here we utilizes the conventional diagonalization method to study such a system in a meticulous process emphasizing the intrinsic property of the system.

When applying the diagonalization methods, we first have to represent the Hamiltonian of the system in a two-dimensional matrix and then diagonalize it for eigenvalues and eigenvectors. The lowest eigenvalues correspond to the bound states of the coupled system, while the eigenvectors can be reorganized to form the wave functions. We first use cubic approximations to diagonalize the Hamiltonian of the system. This scheme is very efficient and can reveal the entanglement of the coupled system. Meanwhile, we extended the Fourier grid method proposed by Marston to two-dimensional problems and found that it also worked here. [20] This diagonalization scheme is based on direct grid point values, which may be thought of as diagonalizing the Hamiltonian on the basis of plane wave functions rather than the basis formed by the eigenstates of a two-dimensional harmonic oscillator. The results are compared with those obtained by imaginary time evolution method. [21]

The rest of the paper includes five sections. In Sect. 2, we give an explicit expression of cubic approximation for the single phase qubit. In Sect. 3, the Hamiltonian for the capacitively coupled phase qubits is derived step by step. Then the first diagonalization scheme based on cubic approximation are performed in Sect. 4 and the other scheme utilizing fast Fourier transform is introduced in Sect. 5. Finally, we give a short summary about the advantages and disadvantages of these methods in Sect. 6.

## 2. Cubic approximation for the single phase qubit

Cubic approximation of the potential of a current-biased single Josephson junction is firstly proposed by Leggett about two decades ago. [22] With critical current $I_c$ and bias current $I_b$, the Hamiltonian of the single qubit can be expressed as

$$\hat{H} = T + V = \frac{\hat{Q}^2}{2C_J} - \frac{I_c \Phi_0}{2\pi}\cos(\hat{q}) - \frac{I_b \Phi_0}{2\pi}\hat{q} \,. \qquad (1)$$

$\hat{Q}$ denotes the charge accumulated around the Josephson junction with $\hat{Q} = 2e\hat{n}$,

where $n$ is the number of corresponding Cooper pairs. $\hat{q}$ is the phase difference across the Josephson junction with $C_J$ its equivalent capacitance. The electrostatic energy of the first part in Eq. (1) is the kinetic part in the Hamiltonian, i.e.,

$$T = \frac{\hat{Q}^2}{2C_J} = \frac{\hat{p}^2}{2m}, \tag{2}$$

where $\hat{p} = \frac{\hbar \hat{Q}}{2e}$ and $m = C_J \left( \frac{\Phi_0}{2\pi} \right)^2$.

The Josephson and bias current energy terms in Eq. (1) are the potential part, and it can be approximated by a polynomial expansion, i. e.,

$$V(q) = -\frac{I_c \Phi_0}{2\pi} \cos(q) - \frac{I_b \Phi_0}{2\pi} q = -E_J [\cos(q) + tq], \tag{3}$$

where $t = \frac{I_b}{I_c}$, $E_J = \frac{I_c \Phi_0}{2\pi}$, and we use $q$ instead of $\hat{q}$. $V(q)$ has two extreme points in the range $[0, \pi]$, the minimum $q_1 = \arcsin(t)$, and the maximum $q_2 = \pi - \arcsin(t)$. Changing the origin of the coordinate axes from $(0,0)$ to $(\arcsin t, V(\arcsin(t)))$ and substituting $q + \arcsin(t)$ for $q$, the new potential can be written as

$$U(q) = E_j [-tq + \sqrt{1-t^2} - \cos(q + \arcsin(t))]. \tag{4}$$

This change is performed because we find that in this situation the potential form can be approximated by polynomials without the zero- and first-order terms, which do not influence the characteristics of the Hamiltonian. Below we will assume this form as the standard form of the potential.

From the configuration of the potential (4), it is better to approximate the nonlinear function using the values at the mininma $(0,0)$ and the turning point $(\arccos(t), U(\arccos(t)))$ because they lie in the middle of the useful potential range and enjoy some specific properties. Set $U_1(q) = Aq^2 - Bq^3$ and obtain the second order derivative at the turning point, we obtain

$$\frac{d^2U_1(q=\arccos(t))}{dq^2}=0\text{, which yields } 2A-6B\arccos(t)=0. \quad (5)$$

Also the second derivatives at the minimum should be equal for $U(q)$ and $U_1(q)$, and it yields $2A=\sqrt{1-t^2}E_J$ . (6)

Results from Eqs. (5) and (6) give the final form of the cubic approximation

$$U(q)\cong U_1(q)=E_J(\frac{1}{2}\sqrt{1-t^2}q^2-\frac{\sqrt{1-t^2}}{6\arccos(t)}q^3). \quad (7)$$

It can be easily seen that the coefficient of the cubic term approximates $-\frac{1}{6}E_J$ as $t\to 1$. The cubic approximation in comparison with the original potential form of Eq. (4) is shown in Fig. 1.

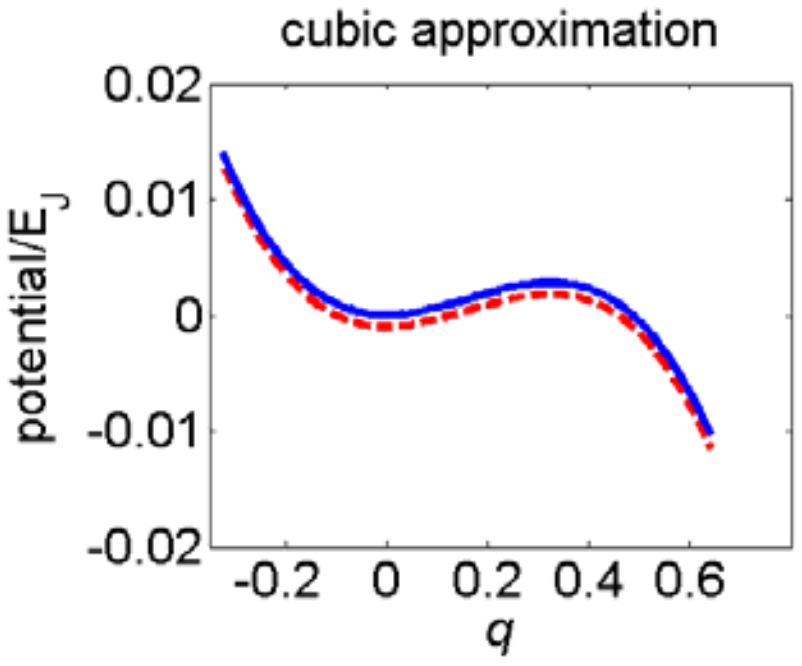

Fig. 1. The blue solid line denotes the original potential, and the red dashed line denotes the cubic approximation shifted downwards by $0.001E_J$. The potential is scaled to $E_J$.

### 3. The Hamiltonian of the capacitively coupled phase qubits

We use a natural process to derive the expression for the Hamiltonian of the coupled system. Starting from the classical form of the Hamiltonian, we can obtain the corresponding quantum mechanical expression. Suppose that the two junctions are fabricated symmetrically with the same Josephson energy scale $E_J$, and the bias constants for each are $t_1$ and $t_2$ respectively. The system is illustrated in Fig. 2.

The potential parts are just the addition of Josephson energies of the two junctions. Using Eq. (4) this part is written as

$$U(q_1,q_2)=E_J[-t_1q_1+\sqrt{1-t_1^{\ 2}}-\cos(q_1+\arcsin(t_1))]$$

$$+\ \ E_J[-t_2q_2+\sqrt{1-t_2^{\ 2}}-\cos(q_2+\arcsin(t_2))]\,. \qquad (8)$$

The electrostatic energy of the coupled system are the sum of the three electrostatic energies for each capacitance. In fact, the coupling term is just coming from the electrostatic energy of the coupling capacitance $C_c$. The sum is just

$$T_{electric}=\frac{1}{2}C_JU_1^{\ 2}+\frac{1}{2}C_JU_2^{\ 2}+\frac{1}{2}C_c(U_1-U_2)^2\,. \qquad (9)$$

With the Josephon relations $U_i=\frac{\hbar}{2e}\dot{q}_i=\frac{\Phi_0}{2\pi}\dot{q}_i$ , the above equation can be transformed to

$$T_{electric}=\frac{1}{2}\left(\frac{\Phi_0}{2\pi}\right)^2(\dot{q}_1,\dot{q}_2)\begin{pmatrix}C_J+C_c & -C_c\\ -C_c & C_J+C_c\end{pmatrix}\begin{pmatrix}\dot{q}_1\\ \dot{q}_2\end{pmatrix}$$

$$=\ \frac{C_J}{2(1-\zeta)}\left(\frac{\Phi_0}{2\pi}\right)^2(\dot{q}_1,\dot{q}_2)\begin{pmatrix}1 & -\zeta\\ -\zeta & 1\end{pmatrix}\begin{pmatrix}\dot{q}_1\\ \dot{q}_2\end{pmatrix}, \qquad (10)$$

where $\zeta=\frac{C_c}{C_J+C_c}$ is defined as the coupling coefficient. Then $C_J+C_c=\frac{1}{1-\zeta}C_J$ and $C_c=\frac{\zeta}{1-\zeta}C_J$ .

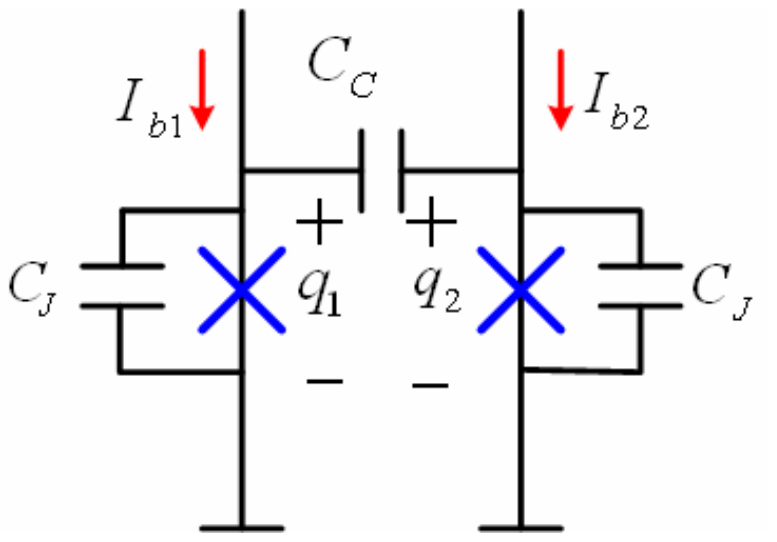

Fig. 2. The capacitively coupled phase qubits**.**

In the situation, the equivalent mass is replaced by a tensor matrix. Suppose that $\vec{p}=\begin{pmatrix}p_1\\ p_2\end{pmatrix}=M\begin{pmatrix}\dot{q}_1\\ \dot{q}_2\end{pmatrix}$, then $\begin{pmatrix}\dot{q}_1\\ \dot{q}_2\end{pmatrix}=M^{-1}\begin{pmatrix}\dot{p}_1\\ \dot{p}_2\end{pmatrix}$. And the kinetic part can be written as $\frac{1}{2}\vec{p}^T\bullet M^{-1}\bullet\vec{p}$ . Considering Eq. (10), we find that

$$\frac{1}{2}\vec{p}^T \bullet M^{-1} \bullet \vec{p} = \frac{1}{2}\left(\frac{\Phi_0}{2\pi}\right)^2 \vec{p}^T \bullet M^{-1} \bullet C \bullet M^{-1} \bullet \vec{p}\,, \tag{11}$$

where $C = \frac{C_J}{1-\zeta}\begin{pmatrix} 1 & -\zeta \\ -\zeta & 1 \end{pmatrix}$.

Eq. (11) yields $M = \frac{C_J}{1-\zeta}\left(\frac{\Phi_0}{2\pi}\right)^2 \begin{pmatrix} 1 & -\zeta \\ -\zeta & 1 \end{pmatrix}$ (12)

and $M^{-1} = \frac{1}{(1+\zeta)C_J}\left(\frac{2\pi}{\Phi_0}\right)^2 \begin{pmatrix} 1 & \zeta \\ \zeta & 1 \end{pmatrix}$. (13)

Therefore the kinetic part can be finally represented by

$$T_{electric} = \frac{1}{2(1+\zeta)C_J}\left(\frac{2\pi}{\Phi_0}\right)^2 (p_1{}^2 + 2\zeta\, p_1 p_2 + p_2{}^2)\,, \tag{14}$$

and its corresponding quantum mechanical formulation is just

$$T = \frac{1}{2(1+\zeta)C_J}\left(\frac{2\pi}{\Phi_0}\right)^2 (\hat{p}_1{}^2 + 2\zeta\, \hat{p}_1 \hat{p}_2 + \hat{p}_2{}^2)\,. \tag{15}$$

**4. Diagonalizing the Hamiltonian using cubic approximation**

We consider the first diagonalization scheme in this section. The total Hamiltonian treated here is as follows:

$$\begin{aligned} H = {} & \frac{1}{2(1+\zeta)C_J}\left(\frac{2\pi}{\Phi_0}\right)^2 (\hat{p}_1{}^2 + 2\zeta\, \hat{p}_1 \hat{p}_2 + \hat{p}_2{}^2) \\ & + E_J\left(\frac{1}{2}\sqrt{1-t_1{}^2}\, q_1{}^2 - \frac{\sqrt{1-t_1{}^2}}{6\arccos(t_1)} q_1{}^3\right) \\ & + E_J\left(\frac{1}{2}\sqrt{1-t_2{}^2}\, q_2{}^2 - \frac{\sqrt{1-t_2{}^2}}{6\arccos(t_2)} q_2{}^3\right). \end{aligned} \tag{16}$$

The coupling term in this expression impedes us from directly approximate the system by a two-dimensional harmonic oscillator, therefore we have to rotate the coordinate axes to annihilate the coupling term in the kinetic part. The transformation should be canonical to satisfy $p_+ q_+ + p_- q_- = 2T$, where $p_+, p_-, q_+, q_-$ are just the corresponding new coordinate variables. The relations are shown below

$$\begin{pmatrix} q_+ \\ q_- \end{pmatrix} = \begin{pmatrix} \frac{1}{\sqrt{2(1+\zeta)}} & \frac{1}{\sqrt{2(1+\zeta)}} \\ \frac{1}{\sqrt{2(1-\zeta)}} & -\frac{1}{\sqrt{2(1-\zeta)}} \end{pmatrix} \begin{pmatrix} q_1 \\ q_2 \end{pmatrix}, \tag{17}$$

and $q_1 = \sqrt{\frac{1+\zeta}{2}} q_+ + \sqrt{\frac{1-\zeta}{2}} q_-$, (18)

$$q_2 = \sqrt{\frac{1+\zeta}{2}} q_+ - \sqrt{\frac{1-\zeta}{2}} q_- . \tag{19}$$ [13)]

Meanwhile $m_1 = m_2 = (1+\zeta) C_{\mathrm{j}} \left( \frac{\Phi_0}{2\pi} \right)^2$, (20)

$$p_+ = m_+ \dot{q}_+ = \sqrt{\frac{1+\zeta}{2}} \left( p_1 + p_2 \right), \tag{21}$$

$$p_- = m_- \dot{q}_- = \sqrt{\frac{1-\zeta}{2}} \left( p_1 - p_2 \right). \tag{22}$$ [14)]

The rotated Hamiltonian can be denoted by

$$H = T(p_+, p_-) + U(q_+, q_-), \tag{23}$$

with $T(p_+, p_-) = \frac{{p_+}^2}{2m} + \frac{{p_-}^2}{2m}$, (24)

and

$$\begin{aligned} U(q_+, q_-) = {} & \frac{1+\zeta}{4} \left( \sqrt{1-{t_1}^2} + \sqrt{1-{t_2}^2} \right) {q_+}^2 + \frac{1-\zeta}{4} \left( \sqrt{1-{t_1}^2} + \sqrt{1-{t_2}^2} \right) {q_-}^2 \\ & + \frac{\sqrt{1-\zeta^2}}{2} \left( \sqrt{1-{t_1}^2} - \sqrt{1-{t_2}^2} \right) q_+ q_- + \left( \frac{1+\zeta}{2} \right)^{\frac{3}{2}} \left( k_1 + k_2 \right) {q_+}^3 \\ & + 3 \frac{1+\zeta}{2} \sqrt{\frac{1-\zeta}{2}} \left( k_1 - k_2 \right) {q_+}^2 q_- + 3 \sqrt{\frac{1+\zeta}{2}} \frac{1-\zeta}{2} \left( k_1 + k_2 \right) q_+ {q_-}^2 \\ & + \left( \frac{1-\zeta}{2} \right)^{\frac{3}{2}} \left( k_1 - k_2 \right) {q_-}^3, \end{aligned} \tag{25}$$

where $k_1 = -\frac{\sqrt{1-{t_1}^2}}{6 \arccos(t_1)}, k_2 = -\frac{\sqrt{1-{t_2}^2}}{6 \arccos(t_2)}$.

The first two terms in (25) are of the second order, and they can be combined with the kinetic part in Eq. (24) to denote a two-dimensional harmonic oscillator with the resonant angular frequencies $\omega_+, \omega_-$, with

$$\omega_+ = \frac{2\pi}{\Phi_0}[\frac{E_J}{2C_J}(\sqrt{1-t_1^2}+\sqrt{1-t_2^2})]^{\frac{1}{2}}, \tag{26}$$

$$\omega_- = \sqrt{\frac{1-\zeta}{1+\zeta}}\omega_+ . \tag{27}$$

Then the Hamiltonian of the coupled system can be expanded in a basis formed by the eigenstates of this 2D harmonic oscillator. We choose the basis as

$\{|m_1, m_2\rangle\langle n_1, n_2|\}$, with the indices 1 and 2 hinting the two dimensions of the 2D oscillator and $m_i, n_i$ =0, 1, 2, … , $N_d$. We denote by $N_d$ the number of eigenstates chosen for diagonalization in each dimension. The matrix for diagonalization has the following form:

$$H_d = \begin{bmatrix} \langle 0,0|H|0,0\rangle & \langle 0,0|H|1,0\rangle & \cdots & \langle 0,0|H|N_d-1,N_d-1\rangle \\ \langle 1,0|H|0,0\rangle & \langle 1,0|H|1,0\rangle & \cdots & \langle 1,0|H|N_d-1,N_d-1\rangle \\ \vdots & \vdots & \ddots & \vdots \\ \langle N_d-1,N_d-1|H|0,0\rangle & \langle N_d-1,N_d-1|H|1,0\rangle & \cdots & \langle N_d-1,N_d-1|H|N_d-1,N_d-1\rangle \end{bmatrix}. \tag{28}$$

The element $\langle m_1, m_2|H|n_1, n_2\rangle$ denotes the following integration,

$$\langle m_1, m_2|H|n_1, n_2\rangle = \iint_{S'(EFGH)} \varphi^*_{m_1}(x_1)\varphi_{n_1}(x_1)H\phi^*_{m_2}(x_2)\phi_{n_2}(x_2)dx_1dx_2 . \tag{29}$$

Here $\varphi_k(x_1)$ represents the wave function referring to $k$-th eigenstate for the first freedom of the 2-D oscillator, and $\phi_l(x_2)$ the $l$-th eigenstate for the second degree of freedom. The matrix elements are computed by numerical integration. Some attention should be paid here. Suppose that the integration area before the rotation is just $-a_1 < q_1 < b_1$, $-a_2 < q_2 < b_2$, where $a_i, b_i$ define the range of the original integration area. Then after the rotation the boundaries become $-c_1 < q_1 < d_1, -c_2 < q_2 < d_2$ (see Fig. 3), with $c_1 = -\frac{a_1+a_2}{\sqrt{2(1+\zeta)}}, d_1 = \frac{b_1+b_2}{\sqrt{2(1+\zeta)}}$, $c_2 = -\frac{a_1+b_2}{\sqrt{2(1-\zeta)}}, d_2 = \frac{a_2+b_1}{\sqrt{2(1-\zeta)}}$.

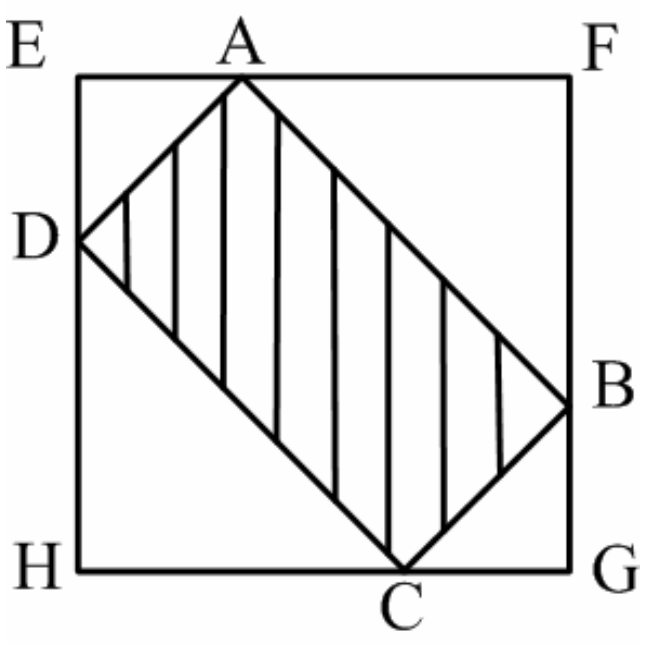

Fig. 3. The integration area for computing the matrix elements.

The area of integration is just the rectangle ABCD, but the area defined by $c_1, d_1, c_2, d_2$ is the square EFGH and marked as $S'(EFGH)$ in the rotated frame. We solve this problem by defining a function over the area EFGH, whose value becomes zero as long as it comes outside the square ABCD. Because the coupling terms or cubic terms in Eq. (25) are expressed in the form of matrix before integration, it is better to perform the definition above after the matrix over area EFGH is obtained.

We choose $N_d = 20$ to diagonalize the Hamiltonian, and the parameters used in the computation are as follows: $t_1 = t_2 = 0.98693$, $I_c = 1.33\times10^{-5} A$, $\zeta = 0.01$.[13] The four lowest states obtained are show in Fig. 4, and the energies are given in the first column of Table 1 marked by Diag-Cubic. The results are compared with those obtained from the imaginary time evolution. The energies are shown in column 3 of Table 1, and the wave functions are of very similar features. Such consistency testifies the validity of this diagonalization scheme.

**Table 1. The energies of the five lowest bound states with three different methods**

| **Eigenenergies(GHz)** | **Diag-Cubic** | **Diag-FFT** | **Imag evolution** |
|---|---|---|---|
| $E_0$ | **6.0585** | **6.0423** | **6.0417** |
| $E_1$ | **11.9449** | **11.8225** | **11.8179** |
| $E_2$ | **12.0032** | **11.8797** | **11.8750** |
| $E_3$ | **17.6533** | **17.1293** | **17.1108** |
| $E_4$ | **17.7935** | **17.1329** | **17.1144** |

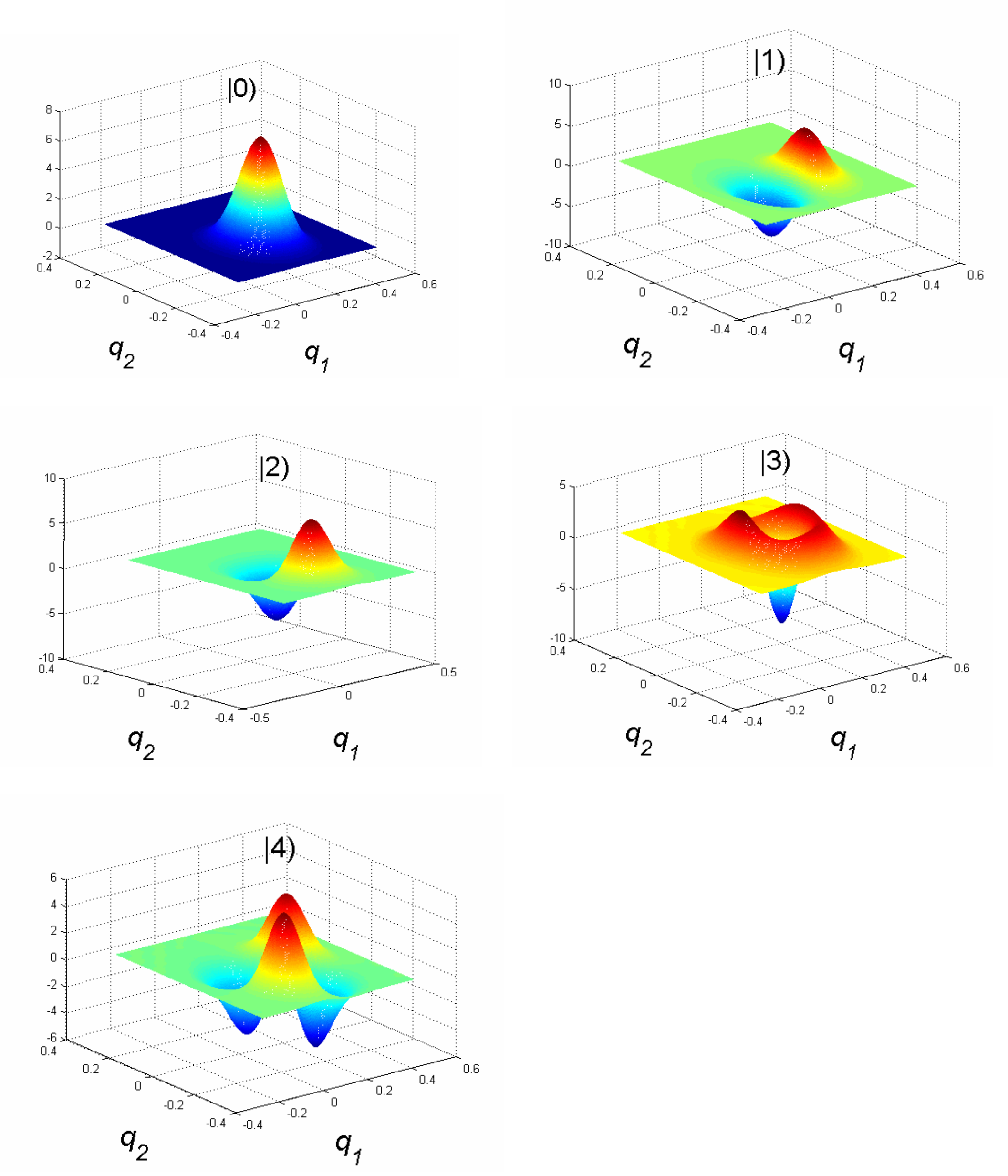

Fig. 4. The wave functions for the lowest five eigenstates of the coupled qubits. They are obtained by the diagonalization method using cubic approximation.

## 5. Diagonalizing the Hamiltonian using Fast Fourier Transform

In this section, Eq. (8) and Eq. (15) are directly employed to denote the Hamiltonian. Because we can operate two-dimensional discrete Fourier transform directly in Matlab programs, there is no need to rotate or compress the axes.

The diagonalization using cubic approximation deals with the kinetic part by absorbing it in the eigen energies of a two-dimensional harmonic oscillator, which in combination with certain second-order terms in the potential just yields $(k-\frac{1}{2})\hbar\omega_{+}+(l-\frac{1}{2})\hbar\omega_{-}$. But here we use discrete Fourier transform to jump from the coordinate space to the kinetic space and then use the inverse discrete Fourier transform to get back to the coordinate space. The units are taken into account although we do not refer much to it. The kinetic part as a whole is represented by a matrix obtained from the above processing.

Suppose $e_{i,j}$ is $N_d \times N_d$ matrix whose values at all points are zeros except the point corresponding to the $i$ th row and the $j$ th column is 1. We perform the inverse two-dimensional FFT to transmit it to the kinetic space, i.e., $s_{i,j} = IFFT2(e_{i,j})$. Meanwhile the kinetic part in Eq. (8) is also represented by $N_d \times N_d$ matrix. Because the bases here are just the plane wave functions, we can simply discretize the form into a matrix $M_k$ without integration. Then the matrix

$$FFT2(M_k \bullet s_{i,j}) = FFT2[M_k \bullet IFFT2(e_{i,j})]$$

denotes one vector of the kinetic part in the coordinate space. We reorganize this matrix into an $N_d{}^2 \times 1$ vector $T_l$ with $l=(j-1)N_d+i$. Repeating the above process, we can finally get an $N_d{}^2 \times N_d{}^2$ matrix $P=\left[T_1, T_2, \cdots, T_{N_d{}^2}\right]$. On the other hand, the potential part in the kinetic space can be directly discretized into an $N_d \times N_d$ matrix $M_u$, and they can be reorganized to a diagonal matrix $V$ where the nonzero elements comes from $M_u$. Finally the matrix form of the Hamiltonian is just $H_f = P + V$. Diagonalizing this matrix will yield the energies and wave functions for the bound states used as computational bases.

It is obvious that this procedure is very universal for problems of $N$-dimension. The main difference lies at the substitution of an $N$-dimension $e_{i_1,i_2,\ldots,i_N}$ for $e_{i,j}$ and the $N$-dimensional fast Fourier transform for the two-dimension fast Fourier transform. Correspondingly, the $N$ -dimensional wave functions should be reorganized from the $N_d^{\,N} \times 1$ vector. The energies for the five lowest eigenstates of this system are also shown in Table 1 marked by Diag-FFT, while the wave functions are similar to that shown in the above section.

## 6. Summary

We have to point out that, in application of both the FFT-diagonalization method and the imaginary time evolution method we have to force periodic boundary conditions while diagonalization method using cubic approximation here set infinite high potential at the boundaries, which may introduce some error. Moreover, the requirement for storage when using diagonalization method increases quickly with increasing $N_d$.

However, as we have shown, both diagonalization methods yield accurate values of the system efficiently and quickly. In particular, the first scheme clearly shows the coupling and entanglement of the system from the vectors after the diagonalization. If we choose $N_d$=5, which means that we choose 5 states for each degree of freedom of the two-dimensional oscillator or approximately for each uncoupled qubit if omitting the cubic terms, we can find the first vector can be reorganized as

$$v_1 = \begin{bmatrix} -0.9923 & -0.0003 & -0.0126 & -0.0001 & -0.0028 \\ -0.1162 & 0.0003 & -0.0260 & 0.0001 & 0.0000 \\ -0.0223 & -0.0002 & -0.0112 & -0.0000 & -0.0023 \\ -0.0172 & 0.0001 & -0.0006 & -0.0000 & -0.0010 \\ -0.0067 & -0.0000 & -0.0018 & 0.0001 & 0.0005 \end{bmatrix}, \qquad (30)$$

The first element (1,1) -0.9923 obviously shows that this state is dominated by the coupling term $|0\rangle|0\rangle$. Also the second vector can be reorganized to $v_2$, with

$$v_2 = \begin{bmatrix} 0.0046 & 0.9713 & -0.0022 & 0.0344 & 0.0001 \\ -0.0386 & 0.2151 & -0.0024 & 0.0453 & -0.0005 \\ -0.0085 & 0.0627 & -0.0012 & 0.0275 & -0.0002 \\ -0.0026 & 0.0231 & -0.0011 & 0.0059 & -0.0000 \\ -0.0013 & 0.0121 & -0.0003 & 0.0025 & -0.0002 \end{bmatrix}, \qquad (31)$$

It is obvious that the first excited state is dominated by the coupling term $|0\rangle|1\rangle$. Such a feature greatly improves our understanding of the physics about a coupling system. On the other hand, the second scheme is very universal and can easily be transplanted to other coupled system with even higher dimensions like that of tripartite qubits.

Stimulating discussions with Genghua Chen, Jilin Wang, Tiefu Li, Peiyi Chen and Zhijian Li are acknowledged. This work is supported by the 211 Program of Nanoelectronics of Tsinghua University.